\documentclass[12pt]{iopart}

\usepackage{graphicx}  
\usepackage{bm}
\begin{document}

\title{Viscous Hydrodynamic Evolution with Non-Boost Invariant Flow for the Color Glass Condensate}

\author{A Monnai$^1$, T Hirano$^{2, 1}$}

\address{$^1$ Department of Physics, The University of Tokyo,
Tokyo 113-0033, Japan}

\address{$^2$ Department of Physics, Sophia University,
Tokyo 102-8554, Japan}

\ead{monnai@nt.phys.s.u-tokyo.ac.jp}

\begin{abstract}
The Large Hadron Collider (LHC) experiments have revealed that the predictions of the color glass condensate (CGC) tend to underestimate the multiplicity at mid-rapidity. 
We develop and estimate a full second-order viscous hydrodynamic model for the longitudinal expansion
to find that the CGC rapidity distributions are visibly deformed during the hydrodynamic stage due to the interplay between the entropy production and the entropy flux to forward rapidity.
The results indicate the importance of viscous hydrodynamic evolution with non-boost invariant flow for understanding the CGC in terms of the heavy ion collisions.
\end{abstract}


\section{Introduction}
The heavy ion program at Large Hadron Collider (LHC) opened up a new era in the physics of the quark-gluon plasma (QGP) \cite{Yagi:2005yb} at higher energies. 
One of the most unique properties of the hot matter is the near-perfect fluidity, which was first discovered in the Au-Au collisions at Relativistic Heavy Ion Collider (RHIC) \cite{Hirano:2008hy}. 
A standard modeling of the heavy ion collision at RHIC consists of several stages; nucleus-nucleus collision, early thermalization, hydrodynamic evolution, freezeout and hadronic cascade. 
The relativistic hydrodynamic model describes the intermediate stage ($\tau \sim$ 1-10 fm/c) where the system is locally equilibrated.
On the other hand, the pre-collision state is considered to be described by the color glass condensate (CGC) where the medium is interpreted as saturated gluons \cite{Gelis:2010nm}. 
The CGC itself is considered to be successful in reproducing the rapidity distributions and the multiplicities observed at RHIC. 
The latest Pb-Pb collisions at LHC, however, revealed that most of the CGC predictions underestimated the multiplicity at mid-rapidity \cite{Aamodt:2010pb}. 
This could be due to the fact that secondary interactions are missing in the comparison. 
In this study we estimate the viscous hydrodynamic evolution of the CGC rapidity distributions at RHIC and LHC with non-boost invariant flow for the first time \cite{Monnai:2011ju}. 
We solve the full second order constitutive equations \cite{Monnai:2010qp} with both shear and bulk viscosity to investigate the hydrodynamic deformation of the initial distributions.

\section{Viscous Hydrodynamic Model}

The viscous hydrodynamic equations consist of the conservation laws and the constitutive equations. 
In the limit of vanishing chemical potentials we consider, the former is the energy-momentum conservation $\partial_\mu T^{\mu \nu} = 0$. We employ the full second-order extended Israel-Stewart theory \cite{Monnai:2010qp, Israel:1979wp} for the latter. 
We choose the Landau frame where the flow $u^\mu$ is identified as the eigenvector of $T^{\mu \nu}$. 
Then the non-equilibrium dynamics in the hydrodynamic system is described by the bulk pressure $\Pi$ and the shear stress tensor $\pi^{\mu \nu}$.
The constitutive equations for the dissipative currents read
\begin{eqnarray}
D \Pi &=& \frac{1}{\tau_{\Pi}} \bigg( -\Pi -\zeta_{\Pi \Pi} \frac{1}{T} \nabla_\mu u^\mu - \zeta_{\Pi \delta e} D \frac{1}{T} \nonumber \\
&+& \chi_{\Pi \Pi}^b \Pi D \frac{1}{T} + \chi_{\Pi \Pi}^c \Pi \nabla_\mu u^\mu + \chi_{\Pi \pi} \pi^{\mu \nu} \nabla_{\langle \mu} u_{\nu \rangle} \bigg) ,
\label{eq:Pi} 
\end{eqnarray}
\begin{eqnarray}
D \pi^{\mu \nu} &=& \frac{1}{\tau_{\pi}} \bigg( -\pi^{\mu \nu} + 2 \eta \nabla^{\langle \mu} u^{\nu \rangle}
+ \chi_{\pi \pi}^b \pi D \frac{1}{T}  
\nonumber \\
&+& \chi_{\pi \pi}^c \pi^{\mu \nu}  \nabla_\rho u^\rho  + \chi_{\pi \pi}^d \pi^{\rho \langle \mu} \nabla_\rho u^{\nu \rangle} + \chi_{\pi \Pi} \Pi \nabla^{\langle \mu} u^{\nu \rangle} \bigg) ,
\label{eq:pi}
\end{eqnarray}
where $D = u^\mu \partial_\mu$ and $\nabla_\mu = \partial_\mu - u_\mu D$ are the time- and the space-like derivatives. 
$T$ is the temperature and $\eta$, $\zeta$, $\tau$ and $\chi$ are the transport coefficients. 
The angle brackets on indices $\langle ... \rangle$ denote the projection of the traceless symmetric components.
We solve the hydrodynamic equations in the (1+1)-dimensional relativistic coordinates to discuss non-boost invariant evolution of the medium, although the effects of the transverse flow are neglected. 
This is in good contrast to the (2+1)-dimensional viscous hydrodynamic models which are used to discuss transverse dynamics but with boost-invariant flow.

One has to introduce the equation of state and the transport coefficients as inputs to perform hydrodynamic calculations. 
Here we employ the equation of state from the latest lattice QCD calculations \cite{Borsanyi:2010cj}. 
Since the transport coefficients are difficult to calculate in the first principle method, the shear viscosity is set to $\eta = s/4\pi$ \cite{Son_visc} and the bulk viscosity $\zeta = \frac{5}{2}(\frac{1}{3}-c_s^2) \eta$ by extending the method in Ref.~\cite{Hosoya:1983id} where $c_s$ is the sound velocity. 
The second order transport coefficients $\tau$ and $\chi$ are calculated with the ratios of the first and the second order ones given in kinetic theory. 
Note that they are employed for the purpose of demonstration to see qualitative responses of the systems. 

The initial energy distributions are calculated from the Monte-Carlo version \cite{Drescher:2006ca, Drescher:2007ax} of Kharzeev-Levin-Nardi (MC-KLN) model \cite{Kharzeev:2001gp, Kharzeev:2002ei}. 
The saturation scale $Q_s$ for a nucleus $A$ at a transverse coordinate $\bm{x}_\perp$ is set as 
\begin{eqnarray}
Q_{s,A}^2(x;\bm{x}_\perp) = 2\ \mathrm{GeV}^2 \frac{T_A (\bm{x}_\perp)}{1.53\ \mathrm{fm}^{-2}}
\bigg( \frac{0.01}{x} \bigg)^\lambda ,
\end{eqnarray}
where $x$ is the momentum fraction of incident partons and $T_A (\bm{x}_\perp)$ the thickness function. 
The parameter $\lambda$ is related with the rapidity dependence of the distribution; $dN_\mathrm{ch}/dy$ gets steeper with increasing $\lambda$. 
Here it is fixed as $\lambda = 0.28$ \cite{Hirano:2009ah}. 
The 5\% most central events are used for the construction of the smooth initial conditions. 
The initial energy distribution $e_0 (\tau_0, \eta_s)$ is obtained from the average CGC transverse energy distribution per unit area $(1/S_\mathrm{area}) dE_T/dy$ by identifying momentum rapidity $y$ with space-time rapidity $\eta_s$. 
Here the initial time is set to $\tau_0 = 1$ fm/c. 
The initial distributions for the dissipative currents are not well-known. 
Here we choose $\Pi  (\tau_0, \eta_s) = 0$ and $\pi^{\mu \nu}  (\tau_0, \eta_s) = 0$ to employ the same initial energy-momentum tensor for ideal and viscous hydrodynamic cases.
The initial flow rapidity is set to $Y_f (\tau_0, \eta_s) = \eta_s$ where $Y_f = \frac{1}{2} \ln \frac{u^0+u^3}{u^0-u^3}$.

We investigate the rapidity distribution through the entropy distribution per flow rapidity $dS/dY_f$, where the two quantities are related as $dN_{\mathrm{ch}}^{\mathrm{hydro}}/dy \approx (2/3)\times (1/3.6) \times dS/dY_{f}$ \cite{Morita:2002av}. 
This is because the ratio of the number density to the entropy density is temperature dependent in the limit of relativistic massless gas, and flow rapidity in a fluid element on average can be identified with momentum rapidity. 
Note that  this is a pure hydrodynamic quantity which does not involve complicated model assumptions such as freezeout and thus one can compare the initial and the final distributions. 

\section{Results}

%
 \begin{figure}[thb]
 \includegraphics[width=0.5\linewidth]{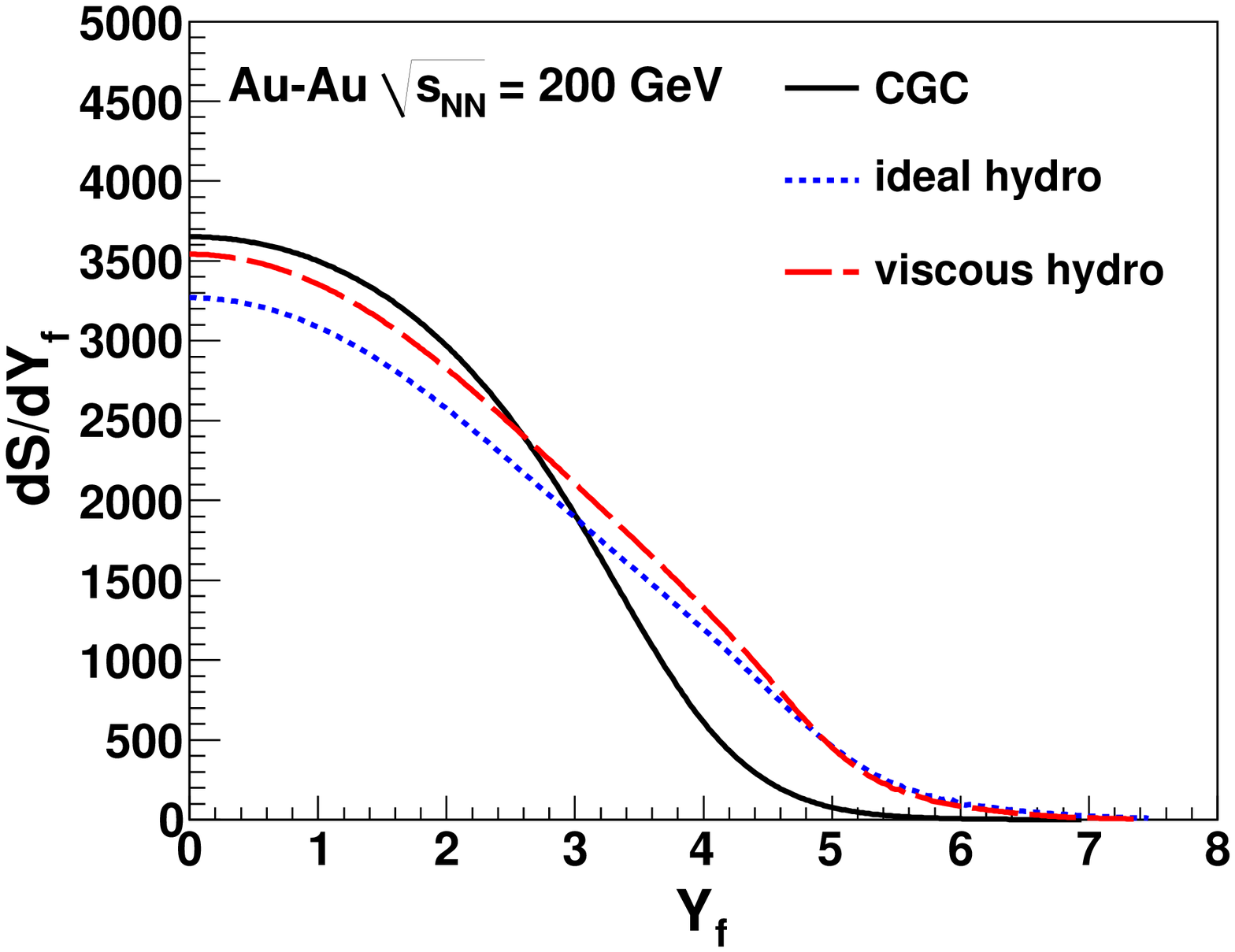}
 \includegraphics[width=0.5\linewidth]{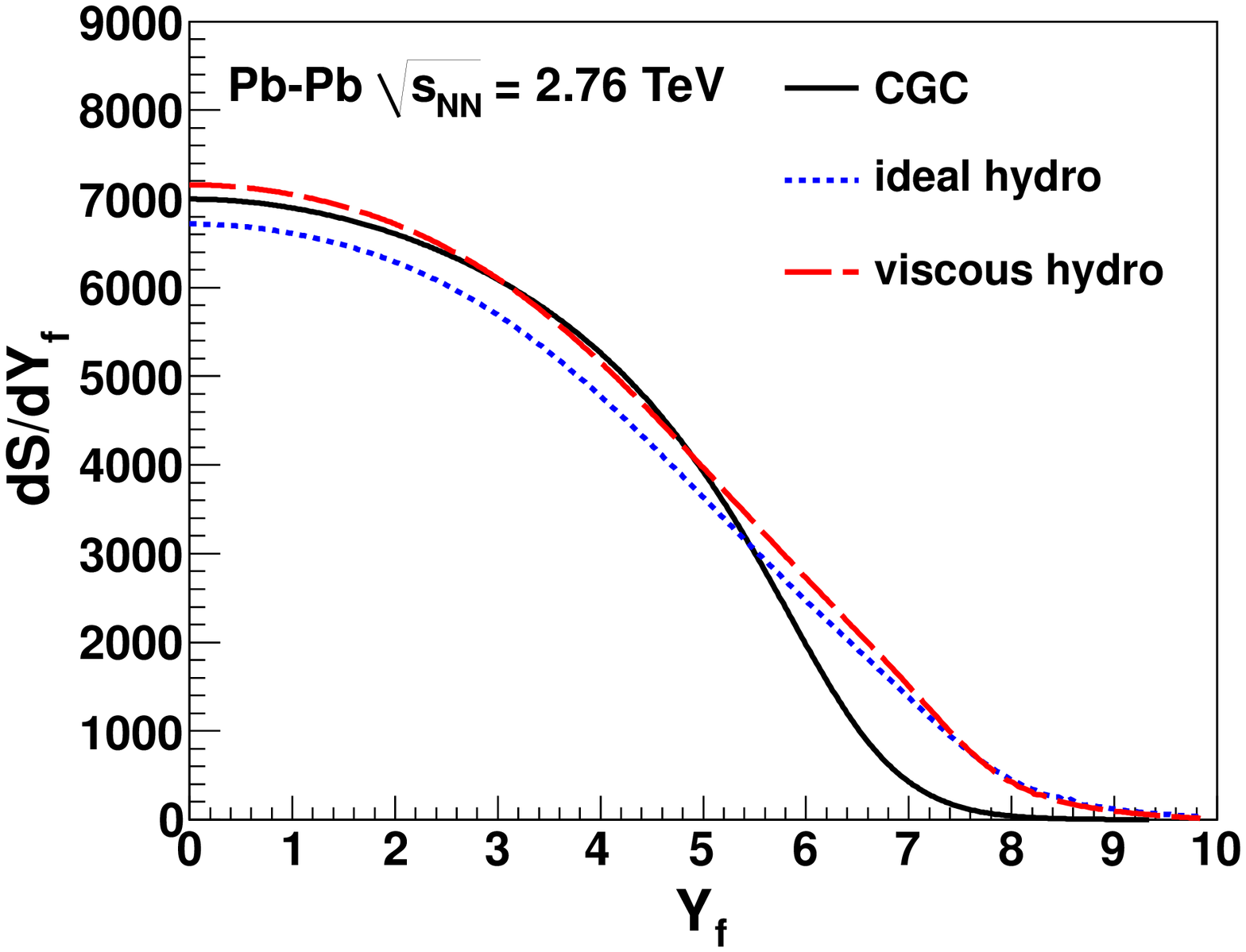}
 \caption{
  The initial rapidity distributions of the CGC (solid lines) modified by ideal (dotted lines) and viscous (dashed lines) hydrodynamic evolution.
  (Left) The RHIC case with the final time $\tau_f = 30$ fm/c.
  (Right) The LHC case with $\tau_f = 50$ fm/c.
 }
 \label{fig:rapiditydist}
 \end{figure}
%

The CGC rapidity distributions with and without hydrodynamic evolutions for Au-Au collisions at $\sqrt{s_{NN}} =$ 200 GeV and Pb-Pb collisions at $\sqrt{s_{NN}} =$ 2.76 TeV are shown in Fig.~\ref{fig:rapiditydist}. 
They correspond to the RHIC and the LHC settings. 
The distributions are always flattened in ideal hydrodynamic systems due to the entropy flux to forward rapidity driven by the pressure gradient in rapidity direction. 
On the other hand, the distributions in viscous hydrodynamic systems are enhanced due to the entropy production. 
The entropy production could be larger because the viscous coefficients here are close to the conjectured lower boundaries. 
The final times are chosen so that the temperatures at mid-rapidity are sufficiently near the QCD pseudo-critical temperature. 
For the current parameter settings, the equal-time surface is not so different than the isothermal one because the distributions do not change much after $\tau \sim 20$ fm/c.

The hydrodynamic modifications differ in magnitude for the RHIC and the LHC cases. 
If, as indicated in the current parameter settings, the CGC rapidity distribution is flattened at RHIC, the actual initial distribution needs to be steeper, \textit{i.e.}, the true $\lambda$ has to be larger than the current value. 
Since the hydrodynamic effects are accidentally cancelled at LHC, it translates into larger multiplicity at LHC. 
This could be one of the candidates to explain the underprediction of the multiplicity at LHC by the CGC. 

\section{Conclusions}

We developed the theoretical and numerical scheme of the full second order viscous hydrodynamic model for the longitudinal expansion. 
The CGC initial rapidity distributions are visibly modified during the hydrodynamic stage due to the interplay between the two factors: (i) entropy production from non-equilibrium processes and (ii) entropy flux to forward rapidity caused by non-boost invariance. 
The result implies that readjustment of the CGC parameters is necessary, which could be one of the key factors for explaining the gap between the CGC predictions and the LHC data \cite{Aamodt:2010pb}. 
The results indicate that non-boost invariant hydrodynamic evolution together with viscosity is indispensable for developing an integrated picture of the heavy ion collisions. 
Detailed analyses on the parameter dependences will also be explored elsewhere \cite{MH}. 

\section*{Acknowledgement}

The authors acknowledge fruitful discussion with Y.~Nara.
The work of A.M. was supported by JSPS Research Fellowships for Young Scientists.
The work of T.H. was partly supported by Grant-in-Aid for Scientific Research No.~22740151.

\end{document}